\documentclass[aip,jcp,11pt,reprint]{revtex4-1}
\usepackage[utf8]{inputenc}
\usepackage{amsmath}
\usepackage{amsfonts}
\usepackage{amssymb}
\usepackage{graphicx}

\begin{document}
\author{F. Bresme} \affiliation{Chemical Physics Section, Department of Chemistry, Imperial College, London SW7 2AZ, United Kingdom and Department of Chemistry, Norwegian University of Science and Technology, Trondheim 7491, Norway}
\author{J.W. Biddle} \affiliation{Institute for Physical Science and Technology, and Department of Chemical and Biomolecular Engineering, University of Maryland, College Park, Maryland 20742, U.S.A.} 
\author{J.V. Sengers} \affiliation{Institute for Physical Science and Technology, and Department of Chemical and Biomolecular Engineering, University of Maryland, College Park, Maryland 20742, U.S.A.}
\author{M.A. Anisimov} \affiliation{Institute for Physical Science and Technology, and Department of Chemical and Biomolecular Engineering, University of Maryland, College Park, Maryland 20742, U.S.A.}

\title{Communication: Minimum in the thermal conductivity of supercooled water: a computer simulation study}
\begin{abstract}
We report the results of a computer simulation study of the thermodynamic properties and the thermal conductivity of supercooled water as a function of pressure and temperature using the TIP4P-2005 water model. The thermodynamic properties can be represented by a two-structure equation of state consistent with the presence of a liquid-liquid critical point in the supercooled region. Our simulations confirm the presence of a minimum in the thermal conductivity, not only at atmospheric pressure, as previously found for the TIP5P water model, but also at elevated pressures. This anomalous behavior of the thermal conductivity of supercooled water appears to be related to the maximum of the isothermal compressibility or the minimum of the speed of sound. However, the magnitudes of the simulated thermal conductivities are sensitive to the water model adopted and appear to be significantly larger than the experimental thermal conductivities of real water at low temperatures. 
\end{abstract}

\maketitle

Supercooled liquid water is known to exhibit anomalous thermodynamic behavior, such as a significant increase in the heat capacity and compressibility with decreasing temperature.\cite{Debenedetti_2003b}  One scenario to account for this observed thermodynamic behavior  assumes a liquid-liquid phase transition terminating in a critical point below the homogeneous nucleation temperature,\cite{Poole_1992,Mishima_1998a,Mishima_1998b} associated with water polyamorphism.\cite{Tanaka_2000,Mishima_2010_PJASB} An equation of state based on a two-structure model is able to represent the thermodynamic properties of real water.\cite{Holten_2012b} While the presence of the liquid-liquid critical point remains a subject of debate,\cite{Limmer_2011,Limmer_2013,Palmer_2013,Overduin_2013,Yagasaki_2014} the thermodynamic behavior of water and water models is consistent with a two-structure equation of state (TSEOS), with or without a phase transition terminating in a critical point.\cite{Holten_2013a,Holten_2014}

This communication is concerned with a possible anomalous behavior of the thermal conductivity of supercooled water. Computer simulations for the TIP5P water model, reported by Kumar and Stanley,\cite{Kumar_2011} indicated that the thermal conductivity of supercooled water displays a minimum as a function  of temperature, in marked contrast to the fluctuation-induced enhancement of the thermal conductivity in the vicinity of the vapor-liquid critical point of water.\cite{IAPWS_ThermalConductivity_2012} In a previous paper,\cite{Biddle_2013} some of us showed that dynamical fluctuation effects are suppressed by the large viscosity of supercooled water. Instead it was suggested that the anomalous behavior of the thermal conductivity $\lambda$ of supercooled water is of thermodynamic origin, possibly through the Bridgman equation:
\begin{equation}
\lambda = 2.8k_Bv^{-2/3}c \label{eq:Bridgman}
\end{equation}
where $k_B$ is Boltzmann’s constant, $v$ the molecular volume and $c$ the speed of sound of the liquid.\cite{Bridgman_1923,Bird_2007}

    To further investigate the anomalous behavior of the thermal conductivity as a function of temperature $T$ and pressure $p$ and its possible relation to the speed of sound, we have performed molecular dynamics simulations of the TIP4P-2005 water model.\cite{Abascal_2005} The TIP4P-2005 model is currently the most accurate non-polarizable water model available \cite{Vega_2011}, and yields adequate representations of densities and compressibilities of real supercooled water \cite{Abascal_2011JCP}, and of the speed of sound in the cold-stable region \cite{Shvab_2013}.  Working with a rigid non-polarizable model enables us to cover long time scales, tens of ns, which are essential to obtain convergent results, particularly at temperatures near the glass transition. Some previous computer simulations have indicated that the TIP4P-2005 model is consistent with the presence of a critical point and a local order consistent with the existence of a high-density and low-density liquid structure.\cite{Wikfeldt_2011a,Yagasaki_2014}

\begin{figure*}
\includegraphics[width=0.99\textwidth]{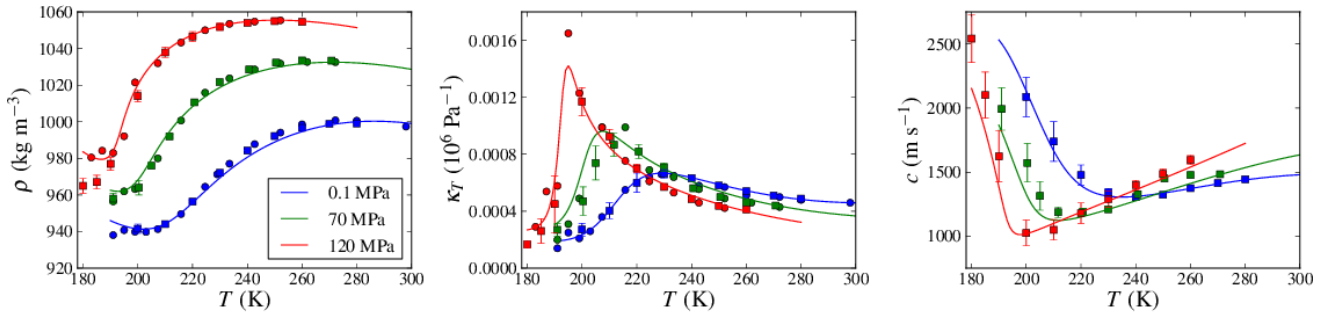}
\caption{Density, compressibility, and speed of sound of TIP4P-2005 at 0.1, 70, and 120~MPa as a function of temperature. The symbols indicate the simulated values: squares (with error bars) obtained in this work and circles (without error bars) obtained previously by Abascal and Vega.\cite{Abascal_2010} The curves represent values calculated from the TSEOS.}
\end{figure*}

    First, the thermodynamic properties at three different pressures, namely at 0.1, 70, and 120~MPa, were determined by performing equilibrium molecular dynamics simulations in the $NpT$ ensemble. We employed cubic simulation boxes with full periodic boundary conditions. A sample consisting of 878 molecules was simulated at each desired pressure and temperature by using the isotropic Parrinello-Rahman barostat~\cite{Parrinello_1981, Nose_1983} and the Nosé-Hoover thermostat.~\cite{Nose_1984,Hoover_1985} The compressibility for the barostat coupling was set to $5\times 10^{-4}$~MPa$^{-1}$. The time constants for the thermostat and the barostat were set to 0.2 ps and 1 ps, respectively, while the equations of motion were integrated with a time step of 2 fs. The molecular interactions were truncated at 1 nm. Long-range corrections for the pressure $p$ and the energy $E$ were included in our computations and the electrostatic interactions were handled with the particle-mesh Ewald method. To obtain convergent results, our simulations covered times from 0.5 to 0.7 $\mathrm{\mu}$s. The equations of state were obtained from equilibrium  simulations performed in parallel with Gromacs 4.5.5.\cite{gromacs_2010} Figure 1 shows the simulated values obtained for the density, isothermal compressibility, and the speed of sound at the three pressures as a function of temperature.  Speed of sound was calculated from heat capacities and compressibilities, which were obtained from analysis of the fluctuations. Our equilibrium properties supplement and agree with previous computations of Abascal and Vega,\cite{Abascal_2010} also shown in Fig. 1 at the pressures considered in this work. We note that TIP4P-2005 in the supercooled regime reaches the diffusive regime at short times ($10^{-8}$~s) compared to our sampling time ($7\times 10^{-7}$~s);  thus we are confident that we obtained equilibrium properties.   We represent the simulated thermodynamic properties by the same type of TSEOS that was previously used by Holten et al. to describe the experimental thermodynamic properties of real water,\cite{Holten_2012b} as well as the properties of the mW and ST2 models.\cite{Holten_2013a,Holten_2014}  The curves in Fig. 1 represent the values calculated from the TSEOS. The TSEOS generally represents the simulated data to within their accuracy, except for some data points at very low temperatures. Our TSEOS implies a critical temperature $T_c = 183\hspace{1mm}\mathrm{K}$, in good agreement with a recent estimates of $T_c = 185\hspace{1mm}\mathrm{K}$ \cite{Yagasaki_2014} and $T_c = 182\hspace{1mm}\mathrm{K}$ \cite{Sumi_2013}.
  
We computed the thermal conductivity of TIP4P-2005 at the same thermodynamic states for which the thermodynamic properties were obtained. All simulations were performed in a microcanonical $NVE$ ensemble with 500 molecules in cubic boxes with full periodic boundary conditions. The electrostatic interactions were computed by using the particle-particle particle-mesh Ewald (PPPM) method with a 1~nm cutoff for the dispersion interactions. A time step of 1~fs was employed for all the thermal-conductivity simulations. The computations were performed with the parallel code LAMMPS.\cite{Plimpton_1995} Equilibrated configurations obtained from the $NpT$ simulations were employed as starting points for the microcanonical simulations. The thermal conductivity was computed with the aid of the Green-Kubo (GK) correlation function:\cite{Zwanzig_1966}
\begin{equation}
\lambda = \frac{V}{3k_BT^2}\int_0^{t_m}dt \left\langle\mathrm{\mathbf{J}}_q(t)\cdot \mathrm{\mathbf{J}}_q(0)\right\rangle, \label{eq:TC}
\end{equation}
\begin{equation}
\mathrm{\mathbf{J}}_q = \frac{1}{V}\left[\sum_i\mathbf{v}_i e_i + \frac{1}{2}\sum_{i\neq j}(\mathbf{f}_{ij}\cdot\mathbf{v}_i)\mathbf{r}_{ij}\right]. \label{eq:Flux}
\end{equation}

\begin{figure}[h!]
\includegraphics[width=0.39\textwidth]{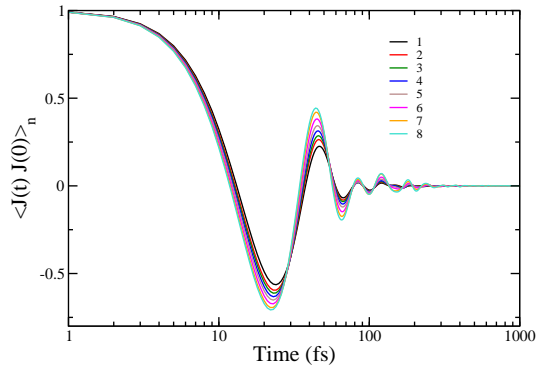}
\caption{Normalized heat-flux correlation functions, $\left< \textbf{\textrm{J}}(t) \cdot \textbf{\textrm{J}}(0) \right>_n = \left< \textbf{\textrm{J}}(t) \cdot \textbf{\textrm{J}}(0)\right>/\left<\textbf{\textrm{J}}(0) \cdot \textbf{\textrm{J}}(0) \right>$, for different systems investigated in this work at $p$=70 MPa. The labels 1 to 8 indicate temperatures: 270.8, 250.5, 240.8, 229.9, 220.7, 211.6 and 191.0 K.}
\end{figure}

In these equations $V$ is the sample volume, $\mathrm{\mathbf{J}}_q$ the heat flux, $e_i$ the energy (kinetic + potential) of atom $i$, $\mathrm{\mathbf{v}}_i$ the velocity of atom $i$, and $\mathbf{f}_{ij}$ the force between atoms $i$ and $j$.  The summations in Eq. (\ref{eq:Flux}) run over all atoms in the system, and include non-bonded and bonded interactions (see ref.~\cite{Sirk_2013}).  The computation of the heat flux with the electrostatic interactions has been discussed previously, both for the Ewald-summation approach\cite{Bresme_1996} and for the PPPM approach\cite{Sirk_2013}. We have chosen the GK method, since it is more effective for resolving the thermal conductivities of thermodynamic states with similar temperatures. The slow dynamics associated with the supercooled states means that the computation of the thermal conductivity requires significant sampling for very long times.  Since the thermal conductivity exhibits a weak dependence on the temperature near a minimum, the simulations at low temperatures needed trajectories of the order of 80~ns. Only with these long time scales were we able to resolve the minima of the thermal conductivity. Averages obtained from short trajectories, e.g., 2~ns, yielded thermal-conductivity values that were too noisy for us to resolve the presence of a minimum. The heat-flux correlation function in the integrand of Eq (\ref{eq:TC}) (Fig. 2) exhibits enhanced oscillations at low temperatures requiring short time steps of 1 fs in the evaluation of the integral. With the choice $t=5~\mathrm{ps}$ for the upper limit, good convergence was found for all the integrations; longer correlation times (up to 10 ps) gave no evidence for decay in the correlation functions.  The results of our simulations of the thermal conductivity for the three pressures are shown in Fig. 3. 

\begin{figure}
\includegraphics[width=0.49\textwidth]{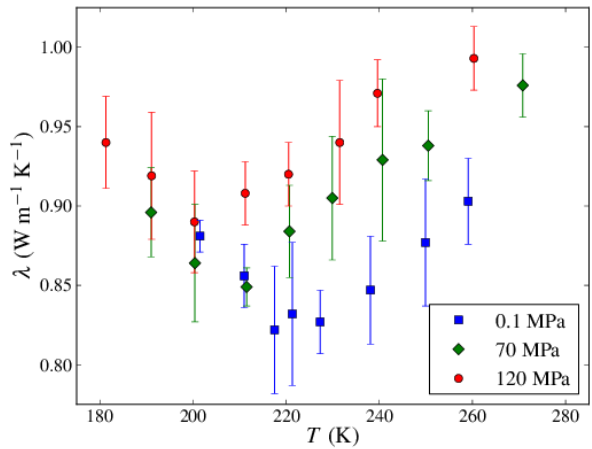}
\caption{Thermal conductivity of TIP4P-2005 at 0.1, 70, and 120~MPa as a function of temperature.}
\end{figure}

We see that the thermal conductivity at each pressure does exhibit a minimum as a function of temperature. The temperature $T_{\mathrm{min}}$, at which the thermal conductivity exhibits a minimum, decreases with increasing pressure. Within computational accuracy, the location of this minimum temperature $T_{\mathrm{min}}$ is correlated with the temperatures of the maximum of the compressibility and with the minimum of the sound velocity, either directly or through the Bridgman equation (\ref{eq:Bridgman}), as shown in Fig. 4. The temperature of maximum compressibility is sometimes referred to in the literature as the Widom temperature.\cite{Kumar_2011,Abascal_2010}

\begin{figure}
\includegraphics[width=0.49\textwidth]{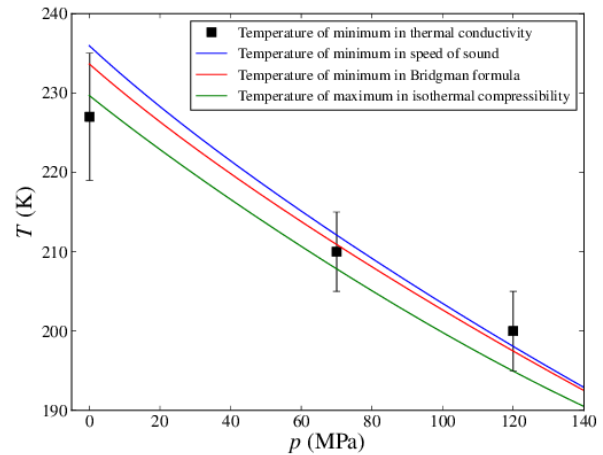}
\caption{Temperature of minimum thermal conductivity compared with the temperature of the maximum of the isothermal compressibility or the minimum of the sound velocity either directly or through the Bridgman equation (\ref{eq:Bridgman}).  Extrema the thermodynamic properties and the Bridgman equation were evaluated using our TSEOS; compressibility maxima agree with Ref. \cite{Abascal_2010}.}
\end{figure}

A simple scale transformation produces a universal curve for the thermal conductivity of the water model in the supercooled state. This is shown in Fig. 5, where we have plotted the thermal conductivity as a function of $T-T_{\textrm{min}}$, while accounting for a small linear dependence of the thermal conductivity on the pressure.   This indicates that the depth of the minimum changes only slowly with pressure, in contrast to the anomaly of the sound velocity.  It also means that unlike the inverse compressibility, which vanishes at the critical point, the thermal conductivity likely remains finite.  Therefore, the thermal conductivity in supercooled water is only partially controlled by thermodynamics. As speed of sound decreases, one should expect other mechanisms of heat transfer, such as particle diffusion, to become more significant.\cite{Rozmanov_2012}

\begin{figure}
\includegraphics[width=0.49\textwidth]{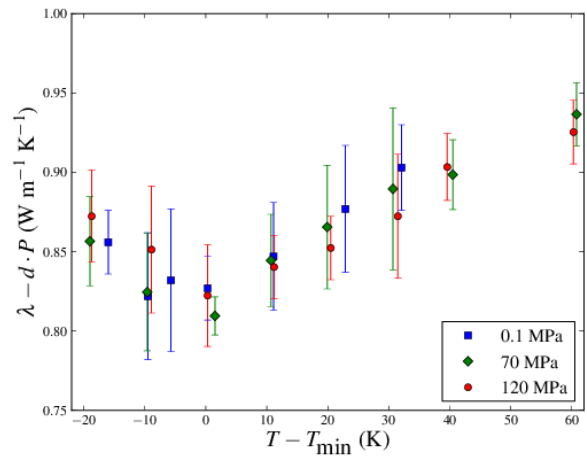}
\caption{Thermal conductivity, corrected for a small linear pressure dependence ($d = 0.564 \hspace{1mm}\mathrm{mW}\hspace{1mm}\mathrm{m}^{-1}\hspace{1mm}\mathrm{K}^{-1}\hspace{1mm}\mathrm{MPa}^{-1}$),  as a function of $T-T_{\mathrm{min}}$.}
\end{figure}

\begin{figure*}[!ht]
\includegraphics[width=0.99\textwidth]{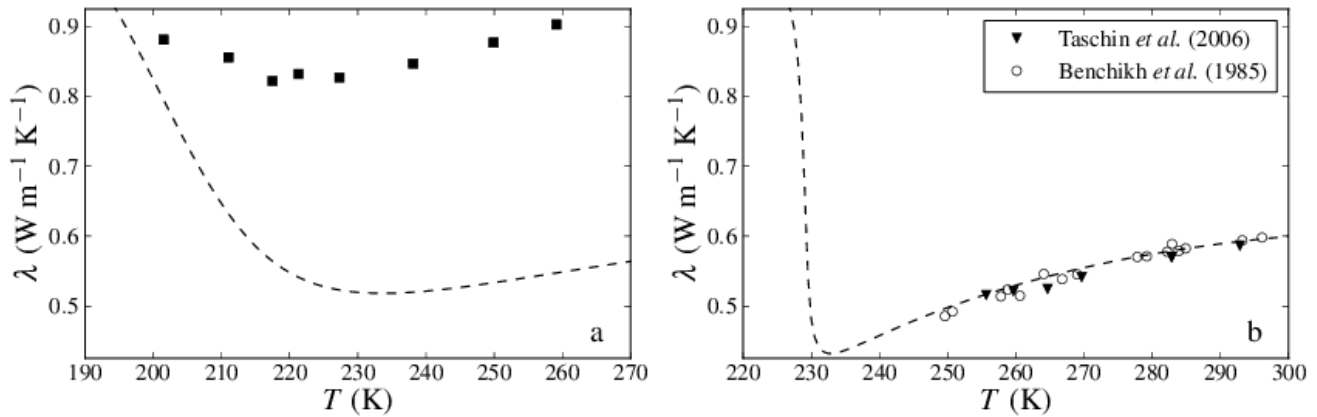}
\caption{Thermal conductivity of TIP4P-2005 at 0.1~MPa (a) and thermal conductivity of real water\cite{Benchikh_1985,Taschin_2006} (b) at 0.1~MPa. The curves represent values calculated from the Bridgman equation (\ref{eq:Bridgman}) for TIP4P-2005 (a) and for real water (b).  The Bridgman formula in each case was evaluated with the TSEOS; in (a) we used our optimization for TIP4P-2005 as elsewhere in this work, while in (b) we used the parameters for real water as presented in Ref. \cite{Holten_2012b}}
\end{figure*}

While our results convincingly demonstrate that the anomalous behavior of the thermal conductivity is of a thermodynamic origin, the values obtained for its magnitude from the simulations are significantly larger that the experimental thermal conductivities of real water. In Fig. 6 we show a comparison between the simulated values of the thermal conductivity (Fig. 6a) and the experimental thermal conductivity data of real water\cite{Benchikh_1985,Taschin_2006} (Fig. 6b) at $p=0.1~\textrm{MPa}$. The simulated thermal conductivities of Kumar and Stanley\cite{Kumar_2011} for TIP5P are even larger ($\sim$ 1.2-1.5 W m$^{-1}$ K$^{-1}$) than those found by us for TIP4P-2005.  The discrepancies between simulated and experimental thermal conductivities are much less at higher temperatures.\cite{Romer_2012}  While the Bridgman equation (\ref{eq:Bridgman}) yields a good quantitative representation of the thermal conductivity of real water, and the Bridgman equation for the model yields values close to the experimental thermal conductivity values in real water, the simulated thermal conductivities of TIP4P are much larger than the values estimated from the Bridgman equation for the model. Hence, it appears that in the supercooled state simulations of thermal conductivity suggest additional heat transport that is not present in real water. A study of the origin of this discrepancy is highly desirable. 

F.B. acknowledges support from EPSRC under Grant EP/J003859/1. F.B. also thanks EPSRC for a Leadership Fellowship. The Imperial College High Performance Computing Service provided computational support. Research of J.W.B. and M.A.A. was supported by the American Chemical Society Petroleum Research Fund under Grant No. 52666-ND6.  

%

\end{document}